\documentclass[summary]{ursi}

\title{Evaluating Direct RF Sampling Performance for RFSoC-based Radio-frequency Astronomy Receivers}
\author{Chao Liu*\affref{ref1}, Larry Ruckman \affref{ref1}and Ryan Herbst \affref{ref1}}

\affiliation{%
  \aff{ref1}{SLAC National Accelerator Laboratory, Menlo Park, CA 94025; e-mail: chaoliu@slac.stanford.edu}
}

\begin{document}

\maketitle

\begin{abstract}

As the maximum RF input and output frequencies of the integrated data converters in RFSoC increase, it becomes practical to digitize and synthesize RF signals in the majority of C band directly without analogue up and down mixing circuits. The elimination of the mixer circuits can significantly simplify the architecture of the receivers or readouts for radio astronomy telescopes. For the systems with large bandwidth or high channel counts, direct sampling can dramatically reduce the size and cost of overall system. This paper with focus on summarising part of the preliminary characterization results for direct sampling with RFSoC data converters in higher order Nyquist zones. 
\end{abstract}

\section{Introduction}

RF system-on-chip (RFSoC) devices have been widely used to develop receivers for radio astronomy applications since it has been released by Xilinx. Some of the applications, such as C-band surveys receivers ~\cite{chao}, the readout for superconducting detectors of microwave SQUID multiplexers (µmux) or microwave kinetic inductance detectors (MKIDs) for Cosmic Microwave Background (CMB) experiments ~\cite{smurf} and millimetre-wavelength telescope after first stage down-conversion ~\cite{chao1}, are operating in the frequency range of 4-8 GHz, which falls in the higher order Nyquist zones of the integrated data converters in RFSoC. Therefore, those receivers requires analogue down-conversion circuits to mix the frequency down to the first Nyqusit zone of the data converters. Due to the high demand of direct RF sampling in telecommunication industry, Xilinx has advanced both the sampling speed and the RF input frequency of the analogue-to-digital converter (ADC) integrated in RFSoC devices. From GEN 1 RFSoC devices to the latest DFE RFSoC devices, the maximum sampling frequency of the ADCs has been increased from 4.096 GHz to 5.9 GHz and the maximum RF input frequency has been extended from 4GHz to 7.125 GHz ~\cite{xilinx}. Those improvements enable us to totally or partially eliminate the analogue down-conversion circuits, which can significantly simplify the architecture of the receiver systems and reduce the hardware cost of the systems, especially for systems with large channel counts. 

The RF signals in higher order Nyquist zone are folded back to the first Nyquist zone, so the RF signal in higher order Nyquist zone can be sampled without down-mixing. Digital-up-coversion (DUC) and digital-down-conversion (DDC) are also included as a part of hardened radio frequency system in RFSoC. The integrated NCOs in DDCs and DUCs can be used to up or down convert the RF signal to desired centre frequency. Therefore, the combination of those components in the hardened radio system in RFSoC can replace the analogue mixing required for some of the applications. In this paper, we present the wide-band performance evaluation results for direct RF sampling schemes with RF data converters and other parts of hardened radio system with different generations of RFSoC devices. The performance of the integrated data converters sampling at first Nyquist zone has been comprehensively discussed in ~\cite{chao}. The focus of this paper is the performance characterization of direct RF sampling at higher orders of Nyquist zones and the integrated DUCs and DDCs. The results can be used as a guide line for future system design and development for RFSoC-based radio-frequency receiver or readout systems with minimum analogue mixing circuits.

\section{Characterization Test Setup}\label{ca}

The characterization described in this paper is performed with the Xilinx Zynq UltraScale+ RFSoC ZCU208 Evaluation Kit, which carries,  XCZU48DR-2FSVG1517E5184 RFSoC, a Gen 3 RFSoc device with eight 14-bit ADCs up to 5 GSPS , and eight 14-bit DACs up to 10 GSPS.

\begin{figure}[htbp]
  \centering
  \includegraphics[width=90mm]{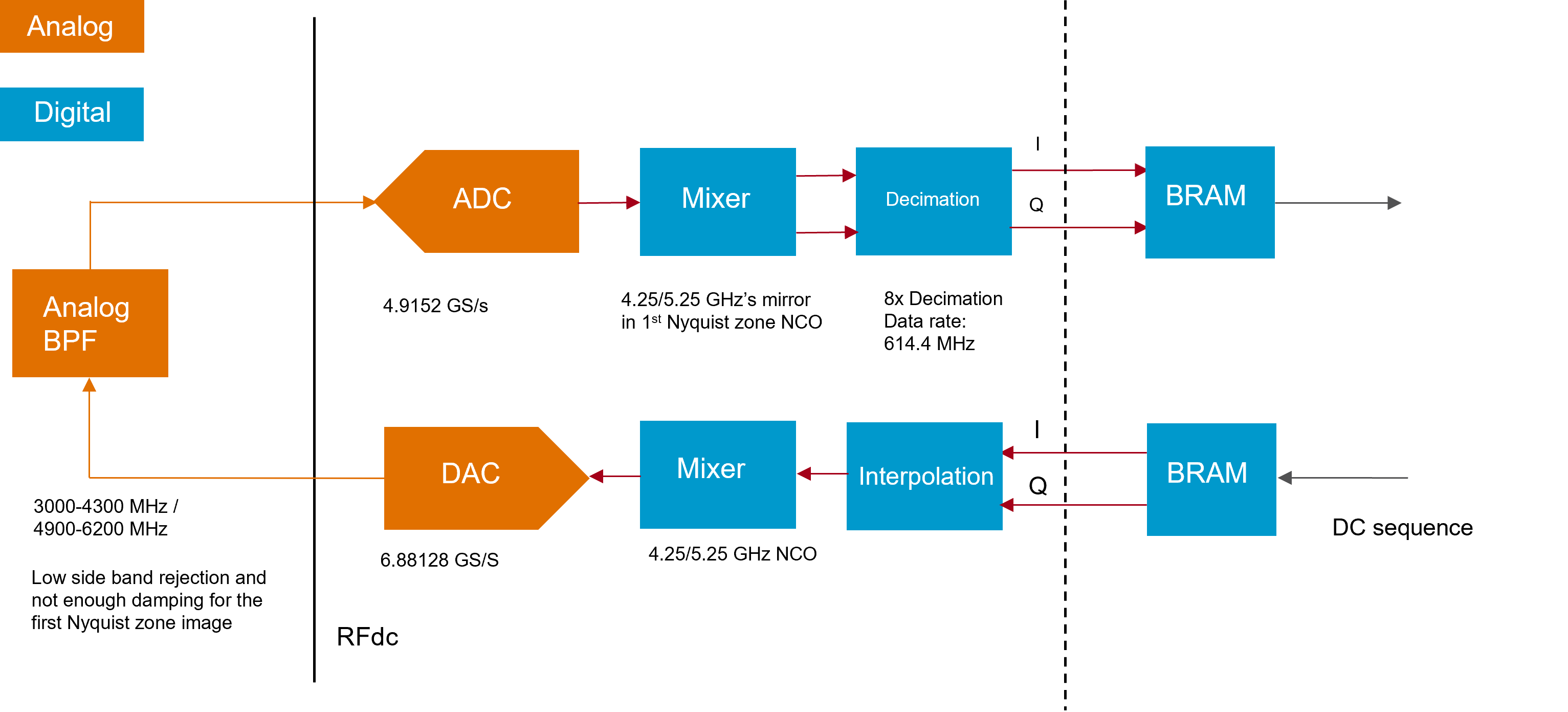}
  \caption{Test setup used for single tone performance characterization. The architecture of data converters remains the same when the DAC is generating a comb of tones, but the configuration of the datapath is changed for different test purposes. }
  \label{fig:testsetup}
\end{figure}

For the targeted applications in this case, both ADCs and DACs are intended to be utilized beyond the first Nyquist zones. Therefore, the first test performed is a single tone test with a loop-back test setup as shown in Figure ~\ref{fig:testsetup}. The RFSoC not only integrate the data converters, but the entire digital up and down conversion datapath, including digital mixers with NCOs and decimation and interpolation hardened blocks for application with narrower bandwidth. 

As Figure ~\ref{fig:testsetup} shows, the characterization is performed with full datapaths on both ADC and DAC side. In this case, the DAC is configured to sample at 6.88128 GHz, which is close to the 7 GHz upper limit when the DAC is in IQ mode. For the single tone test, DC sequency is loaded into the interpolation block in IQ format and the up-mixed digitally at 4.25 or 5.25 GHz, which locates in the second Nyquist zone of DAC. Therefore, the DAC is generating CW RF signal at those two frequencies. The RFSoC DAC has a RF mixed mode to concentrate the RF power in the second Nyquist zone and it has been employed in this test. The output signal of the DAC is filtered with inline band-pass filters to attenuate the images of the RF signal, which can fold back to the first Nyquist zone and appear as spurs in baseband after down conversion.

The ADC sampling speed is 4.9152 GSPS and digital down-mix is performed at the corresponding image frequency of the RF frequency in the first Nyquist zone. The 4.25 GHz RF signal is in the second order Nyquist zone of the ADC and 5.25 GHz in the third order zone. The bandwidth investigated in this case is 600 MHz, Which covers the 500 MHz bandwidth requirement for one of our targeted applications.

\section{Characterization Results}

Two sets of the critical characterizations test results is discussed in this section. The first set is obtained with the exact setup described in Section ~\ref{ca}, which can demonstrate the full spurious free dynamic range (SFDR) for the full loopback circuit. The second setup is performed with similar DAC setup, but the DAC generating a comb of tones and output signal from the DAC is measured by a high frequency spectrum analyzer. 

\subsection{Tests with Single Tones}

\begin{figure}[htbp]
  \centering
  \includegraphics[width=80mm]{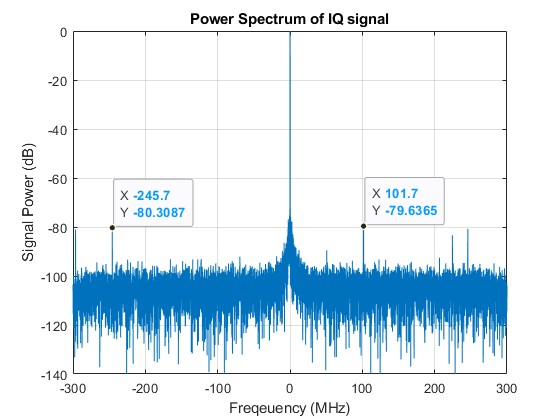}
  \caption{Single tone spectrum with digital up and down conversion at 4.25 GHz or corresponding image frequency at the first Nyquist zone.}
  \label{fig:test425}
\end{figure}

\begin{figure}[htbp]
  \centering
  \includegraphics[width=80mm]{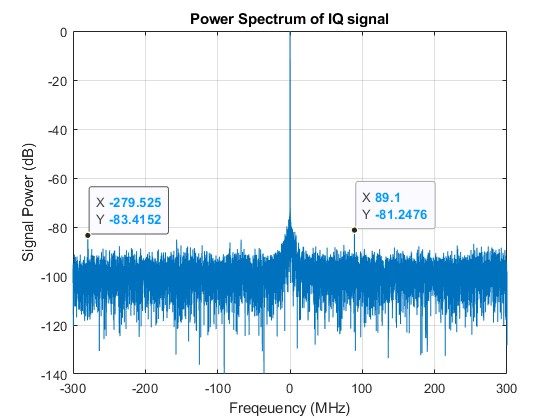}
  \caption{Single tone spectrum with NCOs at 5.25 GHz. The power of the spectrum in normalized the power at primary RF tone.}
  \label{fig:test525}
\end{figure}

The single tone test have been performed at 4.25 GHz and 5.25 GHz. The RF signal has been down-mixed and decimated by the ADC datapath. IQ components are captured at the end of the datapath and the spectrum is calculated offline in Matlab. Figure  ~\ref{fig:test425} and ~\ref{fig:test525} shown the single tone spectrum at 4.25 GHz and 5.25 GHz respectively. As the baseband IQ components are used for spectrum calculation, the 600 MHz bandwidth is centred close to DC, where the tones appear. The SDFR is approximately 79.6 dB at 4.25 GHz and 81.2 dB at 5.25 GHz. The most comparable characterization results listed in Xilinx datasheet ~\cite{xilinx} are the SFDR measurements perform for the ADC of this device family at 4.9 GHz and 5.9 GHz with CW power at –10 dBFS, which are 75 and 74 dB respectively. As the frequency we measured the SFDR are lower than the test cases in datasheet and the test has been performed with signal generator for the results in datasheet, the high SFDR results are reasonable. Therefore, DAC operated at second order Nyquist zone and ADC at second and third orde Nyquist zones can deliver the desired performance for most of the radio astronomy applications~\cite{chao}.

\subsection{Test with Comb of Tones}

\begin{figure}[htbp]
  \centering
  \includegraphics[width=80mm]{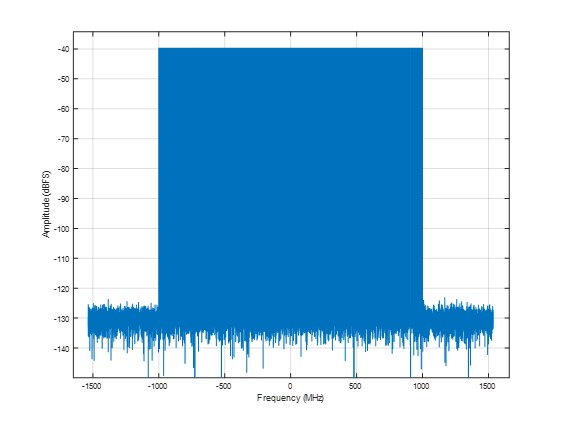}
  \caption{Spectrum of the comb of tones generated in baseband loaded to the integrated DAC in RFSoC.}
  \label{fig:combgen}
\end{figure}

Most of our readout applications employ the frequency division multiplexing (FDM) technique, which requires a large number of tones at different frequencies to be generated by the DAC. The tones are used as probes to measure the phase shift introduced by detector signal, so the phase noise is one of the most critical requirements. In this test, we use one of the integrated DACs to generate a comb of tones as shown in Figure ~\ref{fig:combgen} and the output of the DAC is measured by Keysight EXA signal analyzer N9010B with bandwidth from 10 Hz to 26.5 GHz. As Figure ~\ref{fig:combgen} shown, the tones are generated in baseband from -1  GHz  to 1 GHz in a step size of 2.4 MHz, which covers 2 GHz of bandwidth with 833 tones simultaneously. 

In this test the sampling rate of DAC is 6.144 GHz and the frequency of NCO for up-mixing is 5 GHz. The baseband sequence is generated at 3.072 GHz and then interpolated by a factor of 2.  

\begin{figure}[htbp]
  \centering
  \includegraphics[width=70mm]{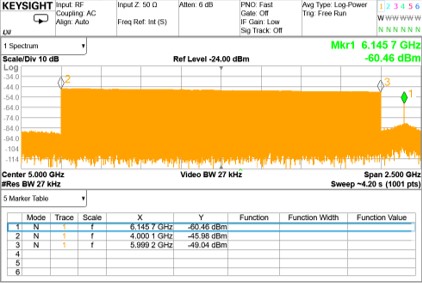}
  \caption{Spectrum of DAC output RF signal captured by the spectrum analyzer.}
  \label{fig:combspec}
\end{figure}

The spectrum of the DAC output captured by the spectrum analyzer is shown in Figure ~\ref{fig:combspec}. The spectrum is centred at 5 GHz and has the 2 GHz bandwidth as expected. The power of the tones reduced by about 3 dB from 4 to 6 GHz, which can be largely attributed to the frequency dependent insertion loss of the balun used for differential to single-ended conversion. The power level per tone around -46 dBm is adequate for the target application and there is no significant intermodulation product or spur in the band of interest. There is a 6.144 GHz spurs in the spectrum, which is the sampling frequency of the DAC and not affecting the performance.  

\begin{figure}[htbp]
  \centering
  \includegraphics[width=70mm]{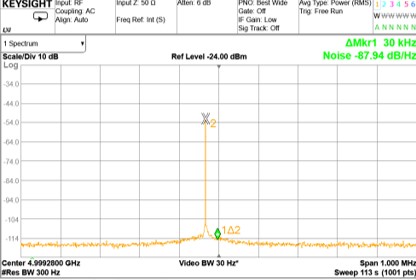}
  \caption{Phase noise measured at 30 kHz offset from one of the tones in the comb.}
  \label{fig:combphase30k}
\end{figure}

\begin{figure}[htbp]
  \centering
  \includegraphics[width=70mm]{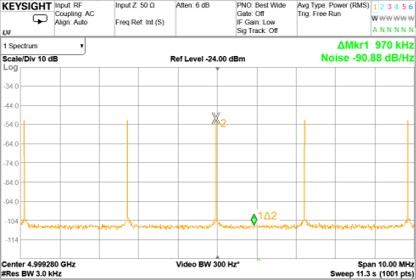}
  \caption{Phase noise measured at about 1 MHz offset from one of the tones in the comb.}
  \label{fig:combphase1m}
\end{figure}

Phase noise measurement is performed with the spectrum analyzer with one of the tones in the comb. The phase noise is measured at 30 kHz and 1 MHz offset with respect to the frequency of the selected tone and the results are shown in Figure ~\ref{fig:combphase30k} and ~\ref{fig:combphase1m}. The phase noise  is about -88 dBc/Hz at 30 kHz offset and -91 dBc/Hz at 1 MHz offset. The results is about 10 dB lower than the system described in ~\cite{smurf}. The system and measurement setup will be investigated and optimized to achieve high phase noise performance. There is no significant intermodulation product in the spectrum given a very high number of tones generated with a single DAC over 2 GHz of bandwidth. 

\textbf{}

\section{Conclusion}

The loopback characterization results demonstrate that the RFSoC integrated data converters and up and down-mixing datapath can deliver required SFDR at higher order Nyquist zones. This can eliminate a large fraction of analogue RF components in the receiver or readout systems for the targeted applications, and therefore, simplify the architecture of system, reduce the cost and offer higher level of flexibility for system configuration. For Gen 3 RFSoC device, the maximum input RF frequency for ADC is 6 GHz, which limits the spectrum converge over C band. However, the direct RF bandwidth has been extended to 7.125 GHz for RFSoC DEF and it can cover higher frequency with analogue mixing.

The test results with a comb of tones shows the DAC's capability of generating a large number of tone with reasonable phase noise and no significant intermodulation noise, which are two of the major concerns for the targeted applications. The characterization results show the great potential of the integrated DAC in RFSoC to be utilized for realizing FDM readouts with high multiplexing factor. As the system development progressing, more comprehensive characterization with optimizations will be performed and published at different stages.

\end{document}